\documentclass[12pt,a4paper]{article}

\usepackage[left=2.5cm,top=2.5cm,bottom=2.5cm,right=2.5cm]{geometry}
\usepackage{graphicx}
\usepackage{amssymb}
\usepackage{amsthm}
\usepackage{amsmath}
\usepackage{amsfonts}
\usepackage{algorithm, algorithmic}
\usepackage{color}

\usepackage{cite}

\usepackage{enumitem}

\newtheorem{thm}{Theorem}[section]
\newtheorem{cor}[thm]{Corollary}

\theoremstyle{definition}

\theoremstyle{remark}
\newtheorem{rem}{Remark}[section]

\numberwithin{equation}{section}

\DeclareMathOperator{\idiv}{\,{\bf div } \,}
\DeclareMathOperator{\imod}{\,{\bf mod } \,}

\begin{document}

\title{The technique of in-place associative sorting}
\author{A. Emre CETIN}

\maketitle

\begin{abstract}

In the first place, a novel, yet straightforward in-place integer value-sorting algorithm is presented. It sorts in linear time using constant amount of additional memory for storing counters and indices beside the input array. The technique is inspired from the principal idea behind one of the ordinal theories of ``serial order in behavior" and explained by the analogy with the three main stages in the formation and retrieval of memory in cognitive neuroscience: (i) practicing, (ii) storage and (iii) retrieval. It is further improved in terms of time complexity as well as specialized for distinct integers, though still improper for rank-sorting.

Afterwards, another novel, yet straightforward technique is introduced which makes this efficient value-sorting technique proper for rank-sorting. Hence, given an array of $n$ {\em elements} each have an integer {\em key}, the technique sorts the elements according to their integer keys in linear time using only constant amount of additional memory. The devised technique is very practical and efficient outperforming bucket sort, distribution counting sort and address calculation sort family of algorithms making it attractive in almost every case even when space is not a critical resource.


\end{abstract}

\begin{description}
\item{keywords:} associative sort, in-situ permutation sort, stimulation sort, linear time sorting.
\end{description}

\section{Introduction}\label{sec:intro}

The adjective ``associative'' derived from two facts where the first one will be realized with the description of the technique. The second one is that, although it replaces all derivatives of the content based sorting algorithms such as distribution counting sort~\cite{Seward,Feurzig}, address calculation sort~\cite{Isaac,Tarter,Flores,Jones,Gupta,Suraweera} and bucket sort~\cite{mahmoud:2000,Cormen} on a RAM, it seems to be more efficient on a``content addressable memory'' (CAM) known as ``associative memory'' which in one word time find a matching segment in tag portion of the word and reaches the remainder of the word~\cite{Hanlon}. In the current version of associative sort developed on a RAM, the nodes of the imaginary subspace (tagged words) and the integers of the array space (untagged words) are processed sequentially which will be a matter of one word time for a CAM to retrieve previous or next tagged or untagged word.

An integer value-sorting algorithm puts an array of {\em integers} into ascending or descending order by their {\em values}, whereas a rank-sorting algorithm puts an array of {\em elements} (satellite information) into ascending or descending order by their numeric {\em keys}, each of which is an integer. It is possible that a rank-sorting algorithm can be used in place of a value-sorting algorithm, since if each element of the array to be sorted is itself an integer and used as the key, then rank-sorting degenerates to value-sorting, but the converse is not always true.

The technique described in this study is suitable for arrays where the elements are laid out in contiguous locations of the memory. Zero-based indexing is considered while accessing the elements, e.g., $A[0]$ and $A[n-1]$ are the first and last elements of the array, respectively, where $n$ is the number of elements of the array.

Nervous system is considered to be closely related and described with the ``serial order in behavior" in cognitive neuroscience~\cite{Lashley,Lashley_1} with three basic theories which cover almost all {\em abstract data types} used in computer science. These are chaining theory, positional theory and ordinal theory~\cite{Henson}.

Chaining theory is the extension of reflex-chaining or response-chaining theory, where each response becomes the stimulus for the next. From an information processing perspective, comparison based sorting algorithms that sort the arrays by making a series of decisions relying on comparing keys can be classified under chaining theory. Each comparison becomes the stimulus for the next. Hence, keys themselves are associated with each other. Some important examples are quick sort~\cite{Hoare}, shell sort~\cite{Shell}, merge sort~\cite{Burnetas} and heap sort~\cite{Williams}. 

Positional theory assumes order is stored by associating each element with its position in the sequence. The order is retrieved by using each position to cue its associated element. Conventional (Von  Neumann) computers store and retrieve order using this method, through routines accessing separate addresses in memory. Content-based sorting algorithms where decisions rely on the contents of the keys can be classified under this theory. Each key is associated with a position depending on its content. Some important examples are distribution counting sort~\cite{Seward,Feurzig}, address calculation sort~\cite{Isaac,Tarter,Flores,Jones,Gupta,Suraweera}, bucket sort\cite{mahmoud:2000, Cormen} and radix sort~\cite{knuth:vol3,mahmoud:2000,sedgewick:algorithms_in_C, Cormen}.

Ordinal theory assumes order is stored along a single dimension, where that order is defined by relative rather than absolute values on that dimension. Order can be retrieved by moving along the dimension in one or the other direction. This theory need not assume either the item-item nor position-item associations of chaining and positional theories respectively\cite{Henson}.

One of the ordinal theories of serial order in behavior is that of Shiffrin and Cook\cite{Shiffrin} which suggests a model for short-term forgetting of item and order information of the brain. It assumes associations between elements and a ``node'', but only the nodes are associated with one another. By moving inwards from nodes representing the start and end of the sequence, the associations between nodes allow the order of items to be reconstructed~\cite{Henson}.

The first technique presented in this study is in-place associative integer sorting~\cite{ecetin,ecetin1,ecetin2,ecetin3}. Inspired from the ordinal model of Shiffrin and Cook, the technique assumes the associations are between the integers in the array space and the nodes in an imaginary linear subspace (ILS) that spans a predefined range of integers. The range of the integers spanned by the ILS is upper bounded by the number of integers $n$ but may be smaller and can be located anywhere provided that its boundaries do not cross over that of the array. This makes the technique in-place, i.e., beside the input array, only a constant amount of memory locations are used for storing counters and indices. An association between an integer of the array space and the ILS is created by a node using a monotone bijective hash function that maps the integers in the predefined interval to the ILS. When a particular distinct integer is mapped to the ILS, a node is created reserving all the bits of the integer except for the most significant bit (MSB) which is used to tag the word as a node of the ILS for interrogation purposes. The reserved bits become the record of  the node which then be used to count (practice) other occurrences of the particular integer that created the node. When all the key of the predefined interval are practiced, the nodes can be stored at the beginning of the array (short-term memory) retaining their relative order together with the information (cue) required to construct the sorted permutation of the practiced interval. Afterwards, the short-term memory is processed and the sorted permutation of the practiced interval is retrieved over the array space in linear time using only constant amount of additional memory. 

Another ordinal theory is the original perturbation model of Estes~\cite{Estes}. Although proposed to provide a reasonable qualitative fit of the forgetting dynamics of the short term memory~\cite{Henson} in cognitive neuroscience, the principle idea behind the method is that the order of the elements is inherent in the cyclic reactivation of the elements, i.e., reactivations lead to reordering of the elements.

In-place associative integer sorting technique is an efficient in-place integer value-sorting algorithm, though not suitable for rank-sorting. Therefore, in-place associative permutation sort~\cite{ecetin4} technique is developed combining the principle idea behind the original perturbation model with the technique of associative integer sorting making it suitable for rank-sorting.


\section{Definitions}\label{sec:pre}

The definition of rank-sorting is: given an {\em array} $S$ of $n$ {\em elements}, $S[0], S[1],\ldots , S[n-1]$ each have an integer {\em key}, the problem is to sort the elements of the array according to their integer keys. To prevent repeating statements like ``key of the element $S[i]$'', $S[i]$ is used to refer the key. 

The definition of value-sorting is: given an {\em array} $S$ of $n$ {\em integers}, $S[0], S[1],\ldots , S[n-1]$, the problem is to sort the integers according to their values. In other words, the elements of the array are integers and used as the keys.

To prevent confusion, the term ``integer'' is used while describing value-sorting techniques, whereas ``key'' is used for rank-sorting.

The notations used throughout the study are: 
\begin{enumerate} [label=({\roman{*}}), nosep]
\item Universe of keys is assumed $\mathbb{U} = [ 0 \ldots 2^{w}-1]$ where $w$ is the fixed word length.

\item Maximum and minimum keys of an array are, $\max (S) = \max(a \vert a \in S)$ and $\min (S) = \min(a \vert a \in S)$, respectively. Hence, range of the keys is, $m = \max (S) - \min (S) + 1$.

\item The notation $B \subset A$ is used to indicated that $B$ is a proper subset of $A$.

\item For two arrays $S_{1}$ and $S_{2}$, $\max (S_{1}) < \min (S_{2})$ implies $S_{1} < S_{2}$.

\end{enumerate}

\section{In-place Associative Integer Sorting}

The most critical phase of associative integer sorting is derived from the cycle leader permutation (in-situ permutation) approach. Hence, a separate section is devised for it.


\subsection{Cycle Leader Permutation}\label{clp}

Given $n$ {\em distinct} integer keys $S[0\ldots n-1]$ each in the range $[\delta, \delta+m-1]$ where $\delta=\min(S)$, if $m=n$ and $S$ is the sorted permutation, then there is a bijective relation $i=S[i]-\delta$ between each key and its position. From contradiction, $i \ne S[i]-\delta$ implies that the key $S[i]$ is not at its exact position. Its exact position can be calculated by $j=S[i]-\delta$. Therefore, the simple monotone bijective hash function $j=S[i]-\delta$ that maps the keys to $j \in [0,n-1]$ can sort the array in $\mathcal{O}(n)$ time using $\mathcal{O}(1)$ constant space. This is cycle leader permutation where $S$ is re-arranged by following the cycles of a permutation $\pi$. First $S[0]$ is sent to its final position $\pi(0)$ (calculated by $j=S[i]-\delta$). Then the element that was in $\pi(0)$ is sent to its final position $\pi(\pi(0))$. The process proceeds in this way until the cycle is closed, that is until the key addressing the first position is found which means that the association $0 = S[0] - \delta$ is constructed between the key and its position. Then the iterator is increased to continue with the key of $S[1]$. At the end, when all the cycles of $S[i]$ for $i=0,1..,n-1$ are processed, all the keys are in their exact position and the association $i = S[i] - \delta$ is constructed between the keys and their position resulting in the sorted permutation of the array. 


\subsection{In-place Associative Distinct Key Sorting}\label{adks}

If we look at the cycle leader permutation closer, we can interpret the technique entirely different. We are indeed creating an imaginary linear subspace $Im[0\ldots n-1]$ over $S[0\ldots n-1]$ where the relative basis of this imaginary linear subspace (ILS) coincides with that of the array space in the physical memory. The ILS spans a predefined interval of the range of keys and this interval is upper bounded by the number of keys $n$. Since the range of the keys $m$ is equal to $n$, it spans the entire range of the keys. The association between the array space and the ILS is created by a node using the monotone bijective hash function $j = S[i] - \delta$ that maps a particular key to the ILS. Since ILS is defined over the array space, mapping a distinct key to the ILS is just an exchange operation from where a node is created. When a node is created for a particular key, the redundancy due to the association between the key and the position of the node releases the word allocated for the key in the physical memory. Hence, we can clear the node and set its tag bit, for instance its most significant bit (MSB) to discriminate it as a node for interrogation purposes, and use the remaining $w-1$ bits of the node for any other purpose. When we want the key back to array space from the ILS, we can use the inverse of hash function and get the key back by $S[i]=i + \delta$ to the array space through the node. However, we don't use free bits of a node for other purposes during cycle leader permutation because it is known that all the keys are distinct and only one key can be mapped to a location creating a node. Therefore, instead of tagging the word as node using its MSB, we use the key itself to tag the word ``implicitly'' as node, since when a key is mapped to the imaginary subspace, it will always satisfy the monotone bijective hash function $i = S[i] - \delta$. Hence, the keys are ``implicitly practiced'' in this case.

This interpretation immediately motivates a rank-sorting algorithm for distinct keys. Consider the problem of sorting $n$ {\em distinct} keys $S[0\ldots n-1]$ each in the range $[\delta, \delta+m-1]$. If $m>n$, it is not possible to construct a monotone bijective hash function that maps all the keys of the array into $j \in [0,n-1]$ without collisions and additional storage space~\cite{Belazzougui}. However, a bijective hash function can be constructed as a partial function that assigns all the keys in the range $[\delta,\delta+n-1]$ to exactly one element in $j \in [0,n-1]$. Hence, a cycle leader permutation only for the keys in the range $[\delta,\delta+n-1]$ with the hash function $j=S[i]-\delta$ can sort these keys in the ILS $Im[0\ldots n-1]$ over $S[0\ldots n-1]$. The hashed keys are said ``implicitly practiced'' and always satisfy the monotone bijective hash function $i = S[i] - \delta$. 

After implicitly practicing the keys in the predefined interval, the next step is to separate the array into practiced and unpracticed keys. This is simply a partitioning problem where we store implicitly practiced keys at the beginning of the array. However, instead of using a pivot for partitioning, we partition the keys that satisfy the hash function $i = S[i] - \delta$. Hence, we obtain a simple two step sorting algorithm for distinct keys:
\begin{enumerate}[label=(\roman{*})., ref=(\roman{*}), itemindent=*]
\item find $\min(S)$ and $\max(S)$ and initialize $\delta = \min(S)$, $\delta' = \max(S)$, $n_d=0$; $n_d'=0$;\label{algo0:item1}
\item implicitly practice all the distinct keys of the interval $[\delta, \delta+n-1]$. We can count the number of implicitly practiced keys in $n_d$ and find the minimum of the unpracticed keys in $\delta'$ ;\label{algo0:item3}
\item store all the implicitly practiced keys (that satisfy $i = S[i] - \delta$) at the beginning of the array;\label{algo0:item4}
\item If $n_d = n$ exit. Otherwise, set $S=S[n_d-1 \ldots n-1]$, $n = n-n_d$, $\delta = \delta'$, $\delta' = \max(S)$, reset $n_d$ and goto step \ref{algo0:item3}.\label{algo0:item5}
\end{enumerate}

The algorithm sorts an array of $n$ elements $S[0\ldots n-1]$ each have a {\em distinct} integer key in the range $[0,m-1]$ using $\mathcal{O}(1)$ extra space in $\mathcal{O}(n+m)$ time for the worst, $\mathcal{O}(m)$ time for the average (uniformly distributed keys) and $\mathcal{O}(n)$ time for the best case. Therefore, the ratio $\frac{m}{n}$ defines the efficiency (time-space trade-offs) letting very large arrays to be sorted in-place. 

The algorithm is surprisingly effective and efficient. Comparisons with $\Omega(n \log n)$ quick sort\cite{Hoare, sedgewick:algorithms_in_C} and merge sort\cite{Anonymous_1} and heap sort\cite{Williams,Levitin} which take $\mathcal{O}(n \log n)$ time on all inputs showed that associative sort for distinct keys is superior than all (up to 20 times) provided that $\frac{m}{n} \le c\log n$ where $c \approx 4$ for heap sort and merge sort. Quick sort gave worser results ($c \approx 8$) for distinct keys. These results are consistent with $m$ calculated theoretically making average case time complexity of the algorithm less than lower-bound of comparison-based sorting algorithms, i.e., $\mathcal{O}(m) < \Omega (n \log n)$. Another very important meaning of this inequality is that, since it does not require additional space other than a constant amount, no matter how large is the array, the proposed algorithm will sort faster than {\em all} provided that $m = \mathcal{O} (n \log n)$.

Comparisons with value-sorting version of distribution counting sort (frequency counting sort~\cite{mahmoud:2000}) showed that associative sort for distinct keys is superior in every case. This is expectable considering memory allocation overload of distribution counting sort since time-complexities of both algorithms are the same. The performance of the algorithm is even {\em asymptotically} better than 2 lines of code referred in textbooks for sorting $n$ distinct integers from $[0\ldots n - 1]$ with indexing an auxiliary array $B$ of the same size as the input $A$ with keys of $A$ by $B[A[i]]=A[i]$ for $i=0,1,\ldots,n-1$, and then reconstructing sorted permutation of $A$ back by $A[i]=B[i]$ for $i=0,1,\ldots,n-1$, which is possibly due to time-consuming memory allocation of the auxiliary array. 

It is compared with radix sort~\cite{knuth:vol3,mahmoud:2000,sedgewick:algorithms_in_C, Cormen} and bucket sort, as well. The results showed that it is superior than radix sort when $\frac{m}{n} \le 8$ and faster than bucket sort for $n$ {\em distinct} integer keys $S[0...n-1]$ each in the range $[0, n-1]$. 

Finally, the dependency of the efficiency of the technique on the distribution of the keys is only $\mathcal{O}(n)$ which means it replaces all the methods based on address calculation~\cite{Isaac,Tarter,Flores,Jones,Gupta,Suraweera}, that are known to be very efficient when the keys have known (usually uniform) distribution and require additional space more or less proportional to $n$~\cite{knuth:vol3}.


\subsection{In-place Associative Integer Sorting}

The technique introduced above that sorts distinct keys can be generalized to arrays with repeating integers if we consider using released $w-1$ bits of a node for other purposes. The generalized version becomes a value-sorting algorithm whereas the former was a rank-sorting algorithm. Henceforth, the term ``integer'' will be used instead of ``key''.

The main difficulties of all distributive sorting algorithms is that, when the integers are distributed using a hash function according to their content, several integers may be clustered around a loci, and several may be mapped to the same location. These problems are solved by inherent three basic phases of in-place associative integer sorting~\cite{ecetin} namely (i) practicing, (ii) storage and (iii) retrieval. 

We will consider the problem of sorting $n$ integers $S[0\ldots n-1]$ each in the range $[\delta, \delta+m-1]$. If $m>n$, it is not possible to construct a monotone bijective hash function that maps all the integers of the array into $j \in [0,n-1]$ without collisions and additional storage space. However, a bijective hash function can be constructed as a partial function that assigns all the integers in the range $[\delta,\delta+n-1]$ to exactly one element in $j \in [0,n-1]$. 


\subsubsection{Practicing}\label{subsec:practicing}

It is assumed that associations are between the integers in the array space and the nodes in an imaginary linear subspace (ILS) that spans a predefined range of integers. The ILS can be defined anywhere on the array space $S[0\ldots n-1]$ provided that its boundaries do not cross over that of the array. The range of the integers spanned by the ILS is upper bounded by the number of integers $n$ revealing the asymptotic power of the technique with increasing $n$ with respect to the range of integers. However, an ILS may be smaller and can be located anywhere over the array space making the technique in-place, i.e., beside the input array, only a constant amount of memory locations are used for storing counters and indices. An association between an integer and the ILS is created by a node using a monotone bijective hash function that maps the integers in the predefined interval to the ILS. Therefore, the monotone bijective hash function is a partial function that assigns to each distinct integer in a predefined interval to exactly one node of the ILS. Since ILS is defined over the array space, mapping a distinct integer to the imaginary subspace is just an exchange operation from where a node is created. This is ``practicing a distinct integer of an interval''. Once a node is created, the redundancy due to the association between the integer and the position of the node (the position where the integer is mapped) releases the word allocated to the integer in the physical memory except for most significant bit (MSB) which tags the word as a node for interrogation purposes. Hence, the integer is said to be sent to the ILS thorugh the node. Nevertheless, the tag bit discriminates the word as a node and the position of the node lets the integer be retrieved back through the node from the ILS using the inverse hash function. This is ``integer retrieval" through the node from ILS. All the bits of the node except the tag bit can be cleared and used to store any information. Hence, they are the ``record'' of the node and the information stored in the record is the ``cue'' by which cognitive neuro-scientists describe the way that the brain recalls the successive items in an order during retrieval. For instance, it will be foreknown from the tag bit that a node has already been created while another occurrence of that particular integer is being practiced providing the opportunity to count other occurrences using the record. The process of counting an other occurrence of a particular integer is ``practicing an idle integer of an interval''. Repeating this process for all the integers of an interval is ``practicing an interval'', i.e., rehearsing used by cognitive neuro-scientists to describe the way that the brain manipulates the sequence before storing in a short (or long) term memory.  Practicing all the integers of an interval does not need to alter the value of other occurrences. Only the first occurrence is altered while being practiced from where a node is created. All other occurrences of that particular integer remain in the array space but become meaningless. That is why they are ``idle integers''. Furthermore, practicing does not need to alter the position of idle integers as well, unless another distinct integer creates a node exactly at the position of an idle integer while being practiced. In such a case, the idle integer is moved to the former position of the integer that creates the new node in its place. This makes associative sort unstable, i.e., equal integers may not retain their original relative order. However, an imaginary subspace can create other subspaces and associations using the idle integers that were already practiced by manipulating either their position or value or both. Hence, a part of linear algebra and related fields of mathematics can be applied on subspaces to solve such problems. 

From information processing perspective, practicing is a derivative of cycle leader permutation (Section~\ref{clp}) where the nodes and the keys that are out of the practiced interval are treated specially. Processing the array from left to right for $i=0,1,\dots n-1$, if $S[i]$ is a node (tagged word) or a key that is out of the practiced interval ($S[i]-\delta \ge n$) increase the iterator $i$ and start over with $S[i]$. Otherwise, it is an integer that is to be practiced ($S[i]- \delta<n$). Therefore, the monotone bijective hash function $j=S[i]-\delta$ maps the keys to $j \in [0,n-1]$. If the integer $S[j]$ at the target is a node, then increase the record of the node by one, increase the iterator $i$ by one and start over with $S[i]$. This is "practicing an idle integer of an interval", because the idle integer $S[i]$ is practiced by its node $S[j]$. On the other hand, if $S[j]$ is not a node, then $S[i]$ is the first occurrence that will create the node. Therefore, copy $S[j]$ over $S[i]$ and then clear $S[j]$ and set its MSB making it a node of ILS releasing its $w-1$ bits free. This is "practicing a distinct integer of an interval" where a node is created with an empty record of $w-1$ bits. In such a case, $j<i$ implies that the integer that was at $S[j]$ (now at $S[i]$) was processed before. Therefore, increase the iterator $i$ by one and start over with $S[i]$. On the other hand, $j \ge i$ implies that the integer that was at $S[j]$ (now at $S[i]$) has not been processed yet. Hence, start over with $S[i]$ without increasing the iterator $i$ to continue with it. 


\subsubsection{Storage}\label{subsec:storage}

Once all the integers in the predefined interval are practiced, the nodes are spread over the ILS depending on the distribution of the integers with relative order. The next step is to store the nodes in a systematic way closing the distance between them to a direction retaining their relative order with respect to each other. This is the storage phase of associative sort where the received, processed and combined information required to construct the sorted permutation of the practiced interval is stored in the short-term memory (beginning of the array). When the nodes are moved, it is not possible to retain the associations between the nodes of the ILS and the integers of the array space because the position of each node cues the recall of the corresponding integer and retrieve it from the ILS using the inverse hash function. This motivates the idea to further use the record of a node to store the node's former position, or maybe its relative position with respect to the ILS or how much that node is moved relative to its absolute or relative position or the other integers. Unfortunately, this requires a record of a particular node is enough to store both the positional information of the node and the number of idle integers practiced by that node. This is statistically impossible, but, as mentioned before, further associations can be created using the idle integers that were already practiced by manipulating either their position or value or both. Hence, if the record is enough, it can store together the positional information and the number of practiced idle integers, whereas an idle integer can be associated accompanying the node to supply additional space to store the positional information if the record is not enough.

Let us assume for a moment that our universe of integers is $\mathbb{U} = [ 0 \ldots 2^{w-1}-1]$ where $w$ is the fixed word length and $n \le 2^{w-1}$. We know that other occurrences of a particular integer can be counted (practiced) using $w-1$ bits (record) of a node. If we decide to write the absolute position of a node into its record during storage, we need $\log n$ bits of the record. Hence, it is logical to think that we can count up to $2^{w-1-\log n}-1$ idle integers with the record of a node during practicing. Fortunately, this is not the case. We can count using all $w-1$ bits of the record during practicing, and while storing the nodes at the beginning of the array (short-term memory), we can get an idle integer immediately after the node that has practiced at least $2^{w-1-\log n}$ idle integers and write the absolute position of that node over the accompanying idle integer. In such a case, the record of the node (predecessor of the idle integer) only stores the number of practiced idle integers. This definition immediately reminds one of the aforementioned main difficulties of all distributive sorting algorithms. When the keys are distributed using a hash function according to their content, several of them may be clustered around a loci. In such a case, how an idle integer can be inserted immediately after a particular node if there are several nodes immediately before and after that particular node during storing? The answer is in the pigeonhole principle. Pigeonhole principle says that,

\begin{cor}
Given $n \le 2^{w-1}$ integers $S[0\ldots n-1]$, the maximum number of distinct integers that may occur contemporary in $S$ at least $2^{w-1-\log n}$ times is,
\begin{equation}
\lceil \frac{n}{2^{w-1-\log n}} \rceil
\end{equation}
\end{cor}
Hence, if the size of the array is say $n = 2^{w-1}$, the maximum number of {\em distinct} integers that may occur contemporary in $S$ at least $1$ time is $n$. But the node itself represents the first occurrence which creates it. Therefore, 
\begin{cor}
The maximum number of nodes that each can practice at least $2^{w-1-\log n}$ idle integers and hence need an idle integer immediately after itself during storage is equal to, 
\begin{equation}\label{eqn:epsilon}
\epsilon = \lceil \frac{n/2}{2^{w-1-\log n}} \rceil
\end{equation}
and upper bounded by $n/2$. 
\end{cor}
This means that, 
\begin{cor}
If the integers that are in the predefined interval $[\delta,\delta+n-\epsilon-1]$ are practiced to $Im[\epsilon \ldots n-1]$ over $S[\epsilon \ldots n-1]$ where $\epsilon \in [0,\frac{n}{2}]$ is calculated by Eqn.\ref{eqn:epsilon}, then there will be $\epsilon$ integers in $S[0 \ldots \epsilon-1]$ either idle or unpracticed (out of the practiced interval) which will prevent collisions while inserting idle integers immediately after the nodes that has practiced at least $2^{w-1-\log n}$ idle integers during storage. 
\end{cor}


\subsubsection{Retrieval}

Finally, the sorted permutation of the practiced interval is constructed in the array space, using the information stored in the short-term memory. This is the retrieval phase of associative sort. It is known that if the record is enough, it stores both the position of the node and the number of practiced idle integers. If not, an associated idle integer accompanying the node stores the position of the node, whereas the record (predecessor of the idle integer) stores the number of practiced idle integers. If the number of occurrences of a particular integer is $n_i$, then there are $n_i-1$ idle integers in the array space. But the nodes represent the first integers that are mapped into the imaginary subspace through themselves. If all the idle integers are grouped on the right side of the short-term memory, then one can process the information in the short-term memory from right to left and distinguish whether there is an idle integer (untagged word) accompanying its predecessor (the node on the left side of the idle integer). An idle integer implies that it is accompanying the node on its left for additional storage. In such a case, the positional information is read from the idle integer, whereas the number of practiced idle integers is read from the record of the node. If there is not an idle integer accompanying the node, both the positional information and the number of practiced idle integers are read from the record of the node. Afterwards, the positional information cues the recall of the integer using the inverse hash function. This is ``integer retrieval'' from ILS. Hence, the retrieved integer can be copied on the array space as many as it occurrs. At this point, we have two options: sequential and recursive versions which will be described next.


\subsubsection{Overall Algorithm}

Consider $n \le 2^{w-1}$ integers $S[0...n-1]$ each in the range $[0,2^{w-1}-1]$. All the integers in the range $[\delta,\delta+n-\epsilon-1]$ will be practiced to $Im[\epsilon, n-1]$ over $S[\epsilon, n-1]$ where $\delta=\min(S)$ and $\epsilon$ is calculated using Eqn.~\ref{eqn:epsilon}. Assume that there are $n_d$ nodes, $n_c$ idle integers practiced by those $n_d$ nodes and $n_d'=n-(n_d+n_c)$ unpracticed integers that are out of the practiced interval. All these can be counted during practicing. Furthermore, the minimum $\delta'$ of the unpracticed integers can be found as well during practicing. 

While $n_d$ nodes are being stored at the beginning of the array (short-term memory) closing the distance between them in order of precedence, if the record of a node is enough, i.e., the node has practiced less than $2^{w-1-\log n}$ idle integers, we write the absolute (former) position of the node into its record together with the number of practiced idle integers. Otherwise, we search to the right and get the first idle integer immediately after the node and write the absolute position of that node over the idle integer. Let us assume that $\epsilon'$ nodes are counted that have practiced at least $2^{w-1-\log n}$ idle integers and needed an idle integer during storage. At the end, $n_d$ nodes and $\epsilon'$ idle integers are stored in the short-term memory $S[0 \ldots n_d+\epsilon'-1]$ with the necessary information required to construct the sorted permutation of the practiced interval. On the other hand, $n_c-\epsilon'$ idle integers and $n_d'=n-(n_d+n_c)$ unpracticed integers are distributed disorderly in $S[n_d+\epsilon' \ldots n-1]$. 


\subsubsection{Sequential Version}

Selecting the pivot equal to $\delta+n-\epsilon-1$, if $n_c-\epsilon'$ idle integers and $n_d'$ unpracticed integers that are distributed disorderly in $S[n_d+\epsilon' \ldots n-1]$ are partitioned, then $n_c-\epsilon'$ idle integers come immediately after the short-term memory resulting in $S[0 \ldots n_d+n_c-1]$. Hence, it is immediate from here that the information in the short-term memory $S[0 \ldots n_d+\epsilon'-1]$ can be processed from right to left backwards and the integers practiced by each node can be copied over $S[0 \ldots n_d+n_c-1]$ right to left backwards without collision with the short-term memory. 


\subsubsection{Recursive Version}

We can recursively practice and store saving $n_d$, $\epsilon'$ and $\delta$ in stack space. Although the exact number of integers to be sorted in the next level of recursion is $n_d'$, the overall number of integers in that recursion is $n=n_d'+n_c-\epsilon'$ where $n_c-\epsilon'$ of them are idle integers of the previous recursion and meaningless. However, these idle integers increase the interval of range of integers spanned by the ILS improving the overall time complexity in each level of recursion. The recursion can continue until no any integer exists. In the last recursion, retrieval phase can begin to construct the sorted permutation of $n_d+n_c$ integers from $n_d+\epsilon'$ records stored in the short term memory $S[0\ldots n_d+\epsilon'-1]$ of that recursion and expand over $S[0 \ldots n-1]$ right to left backwards. Each level of recursion should return the total number of integers copied on the array to the higher level to let it know where it will start to expand its interval. It should be noticed that, in the recursive version of the technique, there is no need to partition $n_c-\epsilon'$ idle integers from $n_d'$ unpracticed integers. Hence, one step is canceled improving the overall efficiency.


\subsubsection{Relaxing the Restrictions}

The technique of associative sorting is explained restricting the universe of integers to $\mathbb{U} = [ 0 \ldots 2^{w-1}-1]$ where $w$ is the fixed word length and $n \le 2^{w-1}$.  

When an integer is first practiced, a node is created releasing $w$ bits of the integer free. One bit is used to tag the word as a node. Hence, it is reasonable to doubt that the tag bit limits the universe of integers because all the integers should be untagged and in the range $[0,2^{w-1}-1]$ before being practiced. Of course we always have the option to use additional $n$ bits to tag the nodes. However, we can,
\begin{enumerate}[label=(\roman{*}), itemindent = * , nosep]
\item partition $S$ into $2$ disjoint sublists $S_1 < 2^{w-1} \le S_2$ in $\mathcal{O}(n)$ time with well known in-place partitioning algorithms as well as in a stable manner with~\cite{Katajainen},
\item shift all the integers of $S_2$ by $-2^{w-1}$, sort $S_1$ and $S_2$ associatively and shift $S_2$ by $2^{w-1}$.
\end{enumerate}
There are other methods to overcome this problem. For instance, 
\begin{enumerate}[label=(\roman{*}), itemindent = * , nosep]
\item sort the sublist $S[0\ldots (n/ \log n)-1]$ using the optimal in-place merge sort~\cite{Salowe},
\item compress $S[0\ldots (n/ \log n)-1]$ by Lemma~1 of~\cite{Franceschini_1} generating $\Omega(n)$ free bits,
\item sort $S[(n/ \log n)\ldots n-1]$ associatively using $\Omega(n)$ free bits as tag bits,
\item uncompress $S[0\ldots (n/ \log n)-1]$ and merge the two sorted sublists in-place in linear time by~\cite{Salowe}.
\end{enumerate}

If practicing a distinct integer lets us to use its $w-1$ bits to practice other occurrences of that particular integer, we have $w-1$ free bits by which we can count up to $2^{w-1}$ occurrences including the node that represents the first integer that created the node. Hence, it is reasonable to doubt again that there is another restriction on the size of the arrays, i.e., $n \le 2^{w-1}$ under the assumption that an integer may always occur more than $2^{w-1}$ times for an array of $n > 2^{w-1}$. But an array can be divided into two parts in $\mathcal{O}(1)$ time and those parts can be merged in-place in linear time by~\cite{Salowe} after sorted associatively.


\subsubsection{Complexity}

From complexity point of view, associative sort shows similar characteristics with distribution counting sort~\cite{Seward,Feurzig} and bucket sort~\cite{mahmoud:2000,Cormen}. It sorts $n$ integers $S[0\ldots n-1]$ each in the range $[0,m-1]$ using $\mathcal{O}(1)$ extra space in $\mathcal{O}(m+n)$ time for the worst, $\mathcal{O}(m)$ time for the average (uniformly distributed integers) and $\mathcal{O}(n)$ time for the best case. The ratio $\frac{m}{n}$ defines the efficiency (time-space trade-offs) of the algorithm letting very larges arrays to be sorted in-place. 

The complexity of the algorithm depends on the number and the range of the integers. It is known that the algorithm is capable of sorting in each iteration (or recursion) the integers in the range $[\delta, \delta+n- \epsilon -1]$ where $\epsilon$ is defined by Eqn.~\ref{eqn:epsilon} and upper bounded by $n/2$. If we restrict our problem to sorting $n$ integers $S[0 \ldots n-1]$ each in the range $[0,n-1]$, the worst case is when $n= 2^{w-1}$ and the complexity is the recursion $T(n) = T(\frac{n}{2}) + \mathcal{O}(n)$ yielding $T(n) = \mathcal{O}(n)$. On the other hand, if we look at Eqn.~\ref{eqn:epsilon} closer, we see that $\epsilon=0$ when $\log n<\frac{w}{2}$. This means that, the best case is when $n < 2^{w/2}$ which implies $\epsilon=0$ and the complexity is exactly $T(n) = \mathcal{O}(n)$ in this case.

For the general case, we should consider the problem of sorting $n$ integers $S[0 \ldots n-1]$ each in the range $[\delta,\delta+m-1]$ where $m>n$. Practically it is always possible that $n-1$ of the integers be in the range $[\delta, \delta+n-\epsilon-1]$. This is the best case and the algorithm sorts the integers in $\mathcal{O}(n)$ time. On the contrary, fixing $\epsilon$ always to its maximum $n/2$, if there is only $1$ integer available in each practiced interval until the last, in any $j$th step, the only integer $s\in S$ that will be sorted satisfies $s < \frac{jn-(j-1)}{2}$ which implies that the last alone integer satisfies $s < \frac{jn-(j-1)}{2} \le m$ from where we can calculate $j$ by $j \le \frac{2m-1 }{n-1}$. In such a case, the time complexity is $\mathcal{O}(n) + \mathcal{O}(n-1) + \dotsc + \mathcal{O}(n-j)<(j+1) \mathcal{O}(n)$ yielding $\mathcal{O}(2m+n)$.

The average case is more difficult to estimate. However, fixing $\epsilon$ always to its maximum $n/2$ will let us to assert that the integers in the range $[\delta,\delta+\frac{n}{2}-1]$ will be sorted in each step. On the other hand, if the integers are uniformly distributed, defining $\beta=\frac{m}{n}$, we can say that $\frac{n}{2\beta}$ integers will be in the range $[\delta,\delta+\frac{n}{2}-1]$ in each step. Therefore, the algorithm will sort $ \frac{n}{2\beta}$ of the integers in $\mathcal{O}(n)$ time in each pass. This will continue until all the integers are sorted and the overall time complexity is,
\begin{equation}\label{eqn:ud_5}
\mathcal{O}(n) \bigl( \frac{(2\beta-1)}{2\beta} + \frac{(2\beta-1)^2}{(2\beta)^2} +\dotsc+ \frac{(2\beta-1)^{k-1}}{(2\beta)^{k-1}} \bigr)
\end{equation}
which means the algorithm is upper bounded by $ 2\beta \mathcal{O}(n)$ or $2\mathcal{O}(m)$ in the average case.


\subsubsection{Empirical Tests}

Practical comparisons for $1$ million 32 bit integers with quick sort showed that associative sort is roughly $2$ times faster for uniformly distributed integers when $m=n$. When $\frac{m}{n} = 10$ performances are same. When $\frac{m}{n} = \frac{1}{10}$ associative sort becomes more than $3$ times faster than quick sort. If the distribution is exponential, associative sort shows better performance up to $\frac{m}{n} \approx 25$ when compared with quick sort. 

Practical comparisons for $1$ million 32 bit integers showed that radix sort is $2$ times faster for uniformly distributed integers when $m=n$. However, associative sort is slightly better than radix sort when $\frac{m}{n} = \frac{1}{10}$. Further decreasing the ratio to $\frac{m}{n} =\frac{1}{100}$, associative sort becomes more than $2$ times faster than radix sort. 

Practical comparisons for $1$ million 32 bit integers showed that associative sort is $2$ times faster than bucket sort with each bucket of size one hence distribution counting sort for $\frac{m}{n} = 1$. Associative sort is still slightly better but the performances get closer when $\frac{m}{n}$ decreases. On the other hand, value-sorting version of distribution counting sort is $2$ times faster than associative sort for $\frac{m}{n} = 1$. Similarly, performances get closer when $\frac{m}{n}$ decreases.  

Even omitting its space efficiency for a moment, associative sort asymptotically outperforms all content based sorting algorithms when $n$ is large relative to $m$. 


\subsection{Improved In-place Associative Integer Sorting}

With a very simple revision, the associative sorting technique can be improved theoretically and practically. The only cost of the improved version is that a recursive implementation is not possible. 

The tag bit discriminates the word as a node in the array space after practicing. During storage where the nodes are stored at the beginning of the array in order of precedence, the positional information ($\log n$ bits) of a node is stored either in its record or in an idle integer accompanying the node. However, ignoring all the MSBs of the array, if only the records ($w-1$ bits) of the nodes are stored at the beginning of the array (short-term memory), there will be unmoved $n_d$ nodes (tagged words) spread over the array space with relative order and $n_d$ records in the short-term memory ($S[0, \ldots n_d-1]$) with the same order after storage. Hence, a one-to-one correspondence is obtained with the stored records and the nodes of the array. This means that, selecting the pivot equal to $\delta+n-1$, if $n_c$ idle integers and $n_d'$ unpracticed integers that are distributed disorderly in $S[n_d \ldots n-1]$ are partitioned, then $n_c$ idle integers come immediately after the short-term memory resulting in $S[0 \ldots n_d+n_c-1]$. Hence, retrieval phase can search the overall array from right to left backwards for the first node, retrieve the integer from the imaginary subspace through that node using its position and inverse of hash function, read its number of occurrences from its record $S[n_d-1]$ and copy over $S[0 \ldots n_d+n_c-1]$ right to left backwards without collision with the short-term memory. Afterwards, the tag bit of the processed node can be cleared and a new search to the left can be carried for the next node which will correspond to the next record $S[n_d-2]$ of the short-term memory. This can continue until all the integers are retrieved resulting in the sorted permutation of the practiced integers.

As the position of the node is not required during storage, there is no need to get an idle integer immediately after a node that has practiced at least $2^{w-1-\log n}$ idle integers. This means that $\epsilon$ is always zero. Hence, the improved version is capable of practicing the integers in the interval $[\delta, \delta+n-1]$ in each iteration. Therefore, while the former technique was capable of sorting integers that satisfy $S[i] - \delta + \epsilon < n$, the improved version sorts the integers that satisfy $S[i] - \delta < n$ in each iteration. Hence, $n$ integers $S[0, \ldots n-1]$ each in the range $[\delta, \delta + n-1]$ can be sorted exactly in $\mathcal{O}(n)$ time regardless of the number of the integers.


\subsection{Improved In-place Associative Distinct Integer Sorting}

The improved associative integer sorting technique can be easily specialized for distinct integers. Given $n$ distinct integers $S[0...n-1]$ each in the range $[\delta, \delta+m-1]$, a monotone bijective {\em super} hash function can be constructed as a partial function that assigns each integer in the range $[\delta,\delta+(w-1)n-1]$ to exactly one element in $j \in [0,n-1]$ and one element in $k \in [0,(w-1)-1]$,
\begin{equation}\label{eqn:shf_1}
j = (S[i] - \delta) \, \idiv \, (w - 1)  \quad  \text{if} \quad S[i] - \delta < (w-1)n  
\end{equation}
\begin{equation}\label{eqn:shf_2}
k = (S[i] - \delta) \imod (w - 1)   \quad \text{if} \quad S[i] - \delta < (w-1)n 
\end{equation}
where $\imod$ is the remainder modulo. In this case, $w - 1$ integers may collide and mapped to the same node created at $j \in [0,n-1]$ (Eqn.~\ref{eqn:shf_1}) in the ILS. But we can use $w- 1$ free bits of a record to encode which of $w-1$ distinct integers are mapped to the same node by setting their corresponding bit determined by $k$ (Eqn.~\ref{eqn:shf_2}). In other words, now the ILS is two dimensional over the array space where the first dimension along the array designates the node position and the second dimension along the bits of the node uniquely determines the integers which are mapped to the ILS through that node.

Once all the distinct integers in the predefined interval are practiced, ignoring all the MSBs of the array, if only the records ($w-1$ bits) of the nodes are stored at the beginning of the array (short-term memory) in order of precedence, there will be unmoved $n_d$ nodes (tagged words) spread over the array space, and $n_d$ records in the short-term memory ($S[0 \ldots n_d-1]$) with the same order after storage. Hence, selecting the pivot equal to $\delta+(w-1)n-1$, if $n_c$ idle integers and $n_d'$ unpracticed integers that are distributed disorderly in $S[n_d \ldots n-1]$ are partitioned, then $n_c$ idle integers come immediately after the short-term memory resulting in $S[0 \ldots n_d+n_c-1]$. Hence, the sorted permutation of the practiced interval can be retrieved by searching the nodes (tagged words) of the array backwards. When a tagged word (node) is found, the base of the integers mapped to that node can be calculated using the position of the node and the inverse of Eqn.\ref{eqn:shf_1}. Then, the position of the bits that are equal to $1$ in the corresponding record $S[n_d-1]$ uniquely determine (with the inverse of Eqn.\ref{eqn:shf_2}) the integers mapped to the same node from most significant to least and each can be copied over $S[0 \ldots n_d+n_c-1]$ right to left backwards without collision with the short-term memory.

As a result, an algorithm is obtained that sorts $n$ distinct integers $S[0 \ldots n-1]$ each in the range $[0, m - 1]$ in exactly $\mathcal{O}(n)$ time if $m < (w-1) n$. Otherwise, it sort in $\mathcal{O}(n+\frac{m}{w-1})$ time for the worst,  $\mathcal{O}(\frac{m}{w-1})$ time for the average (uniformly distributed keys) and $\mathcal{O}(n)$ time for the best case using only $\mathcal{O}(1)$ extra space. 

\section{In-place Associative Permutation Sort}

In-place associative integer sorting introduced in the previous section is a value-sorting algorithm except the one for distinct integer keys (Section~\ref{adks}). All others are not suitable for rank-sorting which is the main objective of all sorting algorithms. In this section, in-place associative permutation technique will be introduced making the former technique suitable for rank-sorting without degrading its performance. 

Although proposed to provide a reasonable qualitative fit of the forgetting dynamics of the short term memory~\cite{Henson} in cognitive neuroscience, the idea behind the original perturbation model of Estes~\cite{Estes} is that the order of the elements is inherent in the cyclic reactivation of the elements, i.e., reactivations lead to reordering of the elements. Hence, when the idea behind the perturbation model is combined with the original technique of associative integer sorting, in-place associative permutation sort is obtained where the order of the practiced interval is inherent in the cyclic reactivation of all the keys of the array. The technique consists of three phases namely (i) practicing, (ii) permutation (reactivation) and (iii) restoring.  

The term ``reactivation'' immediately reminds the cycle leader permutation because when each key is sent to its final position it reactivates the key that was there. Associative permutation sort is based on a very simple idea: if $n_d+n_c$ keys of a predefined interval become consecutive, distinct and in the range $[0, n_d+n_c-1]$, and if each new distinct value of a modified key corresponds to its rank with respect to others pointing its exact position in the sorted permutation, then the overall array can be reactivated with a special form of cycle leader permutation that rearranges the practiced (and modified) keys at the beginning of array in order.

\subsection{Practicing}

Only two new steps are added to practicing phase of associative integer sorting (Section~\ref{subsec:practicing}) before permutation (reactivation) phase. These are (i) accumulation and (ii) re-practicing steps. 

In accumulation step, all the records of the nodes are accumulated from left to right. Hence, at the end, each record of a particular node keeps the exact position of the last idle key practiced by that node. 

As mentioned previously, an ILS can create other subspaces and associations using the idle keys that were already practiced by manipulating either their position or value or both. Hence, it is logical to use the nodes of ILS as discrete hash functions that define the values of idle keys when they are re-practiced using the same monotone bijective hash function. This is the re-practicing step. When an idle key is remapped to its node, it can obtain its exact position (ticket) from the record of its node. The record of the node will be decreased by one for each re-practiced idle key. This means that, when all the idle keys of a particular node are re-practiced, the value of the node will point its exact position in the sorted permutation as well, of course when its MSB is ignored. Hence, when all the idle keys of all the nodes are re-practiced, all the idle keys and all the nodes (ignoring their MSB) will become consecutive, distinct and in the range $[0, n_d+n_c-1]$ pointing to their exact position in the sorted permutation. 

\subsection{Permutation (Reactivation)}

After practicing, if the tag bits of the nodes are ignored, a simple cycle leader permutation can trivially rearrange the practiced keys at the beginning of array in order. Unsurprisingly each node precedes its own idle keys after this rearrangement. This simply puts the elements (satellite information) of the keys in order. However, the modified keys should be restored to their original values unless one intentionally wishes to normalize the keys and put all of them in the range $[0,n-1]$ at the end. Hence, the cycle leader permutation (reactivation) should take care of the nodes since the position of the nodes before reactivation is used to recall the key and retrieve it from the ILS using the inverse hash function. This immediately lets us to assert that if it would be possible to rearrange the practiced interval at the beginning of the array with each node storing its former position in its record, then it would be possible to restore all the modified keys to their original values. A node has a record of $w-1$ bits which stores the node's exact position before the reactivation. Hence, while a node is being moved to its exact position, its former position can be overwritten into its record as the cue which can be used to recall the key using the inverse hash function. But, from information processing perspective, how one can distinguish the nodes that are already moved, from the nodes that are not moved yet in such a case? The idle and unpracticed keys (that are out of the practiced interval) are the solution to this problem. If a node is not at its exact position, then it is evident that either an idle or a key that is out of the practiced interval will address the position of that node. Hence, a special outer cycle leader permutation that only reactivates the idle keys and the keys that are out of the practiced interval will ensure that the corresponding one will be moved to the actual position of the node giving a chance to start an inner cycle leader permutation that reactivates only the nodes and ensures that the nodes will be moved to their exact position storing their former position in their record as the cue. Once a node is moved to its exact position, there can not be any other outer cycle leader which will address that particular position. If another node is available where that particular node is moved, then the inner cycle leader permutation can continue with that node. However, if an idle or an unpracticed key is encountered, then the inner cycle leader permutation terminates and the outer cycle leader permutation continues.

From information processing perspective, associative permutation (reactivation) phase is a derivative of cycle leader permutation (Section~\ref{clp}) where the nodes, practiced idle keys and unpracticed keys (that are out of the practiced interval) are treated specially. Processing the array from left to right for $i=0,1,\dots n-1$, if $S[i]$ is a node (tagged word), do nothing, increase the iterator $i$ and start over with $S[i]$. If $S[i]$ is a key that is out of the practiced interval ($S[i]-\delta \ge n$), exchange $S[i]$ with $S[k]$ starting with $k=n_d+n_c$, and increase $k$ every time a key out of the practiced interval is moved to its new position. Besides, set $j=k-1$ and start over with $S[i]$ if it is either an idle or a key that is out of practiced interval. On the other hand, if $S[i]$ is an idle key ($S[i]-\delta < n$), implicitly practice it by exchanging $S[i]$ with $S[j]$ where $j = S[i]$ and start over with $S[i]$ if it is either an idle or a key that is out of practiced interval. If the key that came to $S[i]$ from $j$ is a node, in other words, if the idle key or the key that is out of the practiced interval is moved to a position where a node is there, then start an inner cycle permutation that reactivates only the nodes until a new idle key or a key that is out of the practiced interval is encountered. To do this, clear MSB of the node $S[i]$ and read where it point by $p=S[i]$. Copy $S[p]$ over $S[i]$ and write the former position $j$ of the node by $S[p]=j$ and set MSB of $S[p]$ making it a node. Now $S[i]$ is the new key to be processed and $j=p$ is where it came from. If $S[i]$ is a node again, continue the cycle leader permutation. However, if $S[i]$ is an idle key or a key that is out of the practiced interval, start over with $S[i]$.

\subsection{Restoring}

With a final scan of the short-term memory ($S[0 \ldots n_d+n_c-1]$), one can obtain the exact values of practiced keys from their preceding nodes (tagged words). Each node stores its absolute position in its record ($w-1$ bits) and cues the recall of the key using the inverse hash function. The value of the key can be copied over all the succeeding keys until a new node is found or the short-term memory ends.

\subsection{Empirical Tests}

Practical comparisons for arrays up to one million integer keys all in the range $[0,n-1]$ on a Pentium machine with radix sort and bucket sort indicate that associative permutation sort is slower roughly $2$ times than radix sort and slower roughly $3$ times than bucket sort. On the other hand, it is faster than quick sort for the same arrays roughly $1.5$ times.
 
Although its time complexity is similar to that of in-place associative sort~\cite{ecetin} and practically slower, in-place associative permutation sort is proposed for {\em integer key sorting} problem.  Hence, it sorts $n$ records $S[0\ldots n-1]$ each have an integer key in the range $[0,m-1]$ with $m>n$ using $\mathcal{O}(1)$ extra space in $\mathcal{O}(n+m)$ time for the worst, $\mathcal{O}(m)$ time for the average (uniformly distributed keys) and $\mathcal{O}(n)$ time for the best case.


\begin{rem}
Associative permutation sort technique is on-line in a manner that after each iteration, the keys in the range $[\delta, \delta+ n-1]$ are sorted at the beginning of the array and ready to be used.
\end{rem}

\section{Conclusions}
\label{chap:summaryandconclusion}

The technique of associative sorting is presented.  In-place associative permutation sort technique is proposed which solves the main difficulties of distributive sorting algorithms by its inherent three basic steps namely (i) practicing, (ii) permutation and (iii) restoring. It is very simple and straightforward and around 30 lines of C code is enough. 

From time complexity point of view, both techniques show similar characteristics with bucket sort and distribution counting sort. They sorts the integers/keys associatively in $\mathcal{O}(m)$ time for the average (uniformly distributed keys) and $\mathcal{O}(n)$ time for the best case. Although its worst case time complexity is $\mathcal{O}(n+m)$, it is upper bounded by $\mathcal{O}(n^2)$ for arrays where $m>n^2$. On the other hand, it requires only $\mathcal{O}(1)$ additional space, making it time-space efficient compared to bucket sort and distribution counting sort. The ratio $\frac{m}{n}$ defines its efficiency (time-space trade-offs) letting very large arrays to be sorted in-place. Furthermore, the dependency of the efficiency on the distribution of the keys is $\mathcal{O}(n)$ which means it replaces all the methods based on address calculation, that are known to be very efficient when the keys have known (usually uniform) distribution and require additional space more or less proportional to $n$. Hence, in-place associative permutation sort asymptotically outperforms all content based sorting algorithms making them attractive in almost every case even when space is not a critical resource.

The technique seems to be very flexible, efficient and applicable for other problems as well, such as hashing, searching, succinct data structures, gaining space, etc. 

The only drawback of associative permutation sort is that it is unstable. However, as mentioned before, an imaginary subspace can create other subspaces and associations using the idle integers that were already practiced by manipulating either their position or value or both. Hence, different approaches can be developed to solve problems such as stability. 

\section{Discussion}
\label{chap:discussion}

Associative permutation sort first finds the minimum of the array and starts with the keys in $[\min(S), \min(S)+n-1]$. However, instead of starting with this interval, omitting the MSBs, if we consider a word as $w-1$ bits and the most significant $\lceil \log n \rceil$ bits of a word as the key and the remaining bits as the satellite information, the problem reduces to sorting $n$ integer keys $S[0 \ldots n-1]$ each in the range $[0,2^{\lceil \log n \rceil}-1]$. Since it is possible that $2^{\lceil \log n \rceil} -1 > n-1$, the keys in $[n, 2^{\lceil \log n \rceil}-1]$ become the keys that are out of the practiced interval.


As a result, when the keys are sorted according to their most significant $\lceil \log n \rceil$ bits, in-place associative most significant $\lceil \log n \rceil$ radix sort is obtained. After the array is sorted according to their most significant $\lceil \log n \rceil$ bits, the idle keys are grouped and each group is preceded by the corresponding node that has practiced them. Hence, each group can be sorted sequentially or recursively assuming the satellite information as the key. If itself is used, it becomes an algorithm based on {\em hash-and-conquer} paradigm in contrast to {\em divide-and-conquer}. However, size of subgroups decreases and it may not be efficient when the ratio of range of keys in each subgroup to size of that subgroup, i.e., $\frac{m}{n}$ increases. Hence, other strategies may need to be developed after the first pass.





\begin{thebibliography}{}

\bibitem{ecetin} {A.E. Cetin, ``In-place associative integer sorting'', arXiv:1209.0572v2 [cs.DS]}
\bibitem{ecetin1} {A.E. Cetin, ``Improved in-place associative integer sorting'', arXiv:1209.3668v1 [cs.DS]}
\bibitem{ecetin2} {A.E. Cetin, ``Sorting distinct integer keys using in-place associative sort'', arXiv:1209.1942v2 [cs.DS]}
\bibitem{ecetin3} {A.E. Cetin, ``Sorting distinct integers using improved in-place associative sort'', arXiv:1209.4714v1 [cs.DS]}
\bibitem{ecetin4} {A.E. Cetin, ``In-place associative permutation sort'', arXiv:arXiv:1210.1771v1 [cs.DS]}

\bibitem{Lashley} {K.S. Lashley, ``The problem of serial order in behavior'', in Cerebral Mechanisms in Behavior, ed. LA Jeffress, John Wiley \& Sons, 1966.}
\bibitem{Lashley_1} {K.S. Lashley, ``In search of the engram'', IEEE Transactions on Electronic Computer, Vol. EC-15, no. 4, 1966.}
\bibitem{Henson} {R.N.A. Henson, ``Short-term memory for serial order: The start-end model'', Cognitive Psychology, Vol. 36, pp. 73~-~137, 1998.}

\bibitem{Hoare}{C.A.R Hoare, ``Quicksort'', Comput. J., Vol. 5, pp. 10~-~16, 1962.}
\bibitem{Shell}{D.L. Shell, ``A High Speed Sorting Procedure'', Communications of ACM, Vol. 2, pp. 30~-~32, 1959.}
\bibitem{Burnetas}{A. Burnetas, D. Solow, R. Agrawal, ``An analysis and implementation of an efficient in-place bucket sort'', Acta  Informatica, Vol. 34, pp. 687~-~700, 1997.}

\bibitem{Williams} {J. Williams, ``Heapsort'', Communications of the ACM, Vol. 7, pp. 347~-~348.}
\bibitem{Anonymous_1} {Anonymous, ``Merge Sort'', Wikipedia, 2012.}
\bibitem{Levitin}{A. Levitin, The Design and Analysis of Algorithms. Addison-Wesley, 2007.}


\bibitem{Seward} {H.H. Seward, Information Sorting in the Application of Electronic Digital Computers to Business Operations, Master's thesis, MIT Digital Computer Laboratory, Report R-232, Cambridge, 1954.}

\bibitem{Feurzig} {W. Feurzig, ``Algorithm 23, mathsort'', Commun. ACM, Vol. 3, pp. 601~-~602, 1960.}


\bibitem{Isaac} {E.J. Isaac, R.C. Singleton, ``Sorting by address calculation'', Journal of the ACM,  Vol. 3, pp. 169~-~174, 1956.}

\bibitem{Tarter} {M.E. Tarter, R.A. Kronmal, ``Non-uniform key distribution and address calculation sorting'', Proc. ACM Nat'l Conf. 21, 1966.}
\bibitem{Flores} {I. Flores, ``Computer time for address calculation sorting'', Journal of the ACM, Vol. 7, pp. 389~-~409, 1960.}
\bibitem{Jones} {B. Jones, ``A variation on sorting by address calculation'', Communications of the ACM , Vol. 13, pp. 105~-~107, 1970.}

\bibitem{Gupta} {G. Gupta, ``Sorting by hashing and inserting'', Proc. ACM Annual Computer Science Conf. 17, pp. 409~-~409, 1989.}
\bibitem{Suraweera} {F. Suraweera, J.M. Al-Anzy, ``Analysis of a modified address calculation sorting algorithm'', Comput. J. Vol. 31, pp. 561~-~563, 1988.}

\bibitem{mahmoud:2000} {H.M. Mahmoud, Sorting, A Distribution Theory, John Wiley and Sons, 2000.}
\bibitem{Cormen} {T.H. Cormen, C.E. Leiserson, R.L. Rivest, C. Stein, Introduction to Algorithms, The MIT Press, 2001.}
\bibitem{sedgewick:algorithms_in_C}{R. Sedgewick, Algorithms in C, Parts 1-4: Fundamentals, Data Structures, Sorting, Searching, Addison-Wesley, 1997.}
\bibitem{knuth:vol3} {D.E. Knuth, The Art of Computer Programming, Volume 3: Sorting and Searching, Addison-Wesley, 1973.}

\bibitem{Shiffrin} {R. Shiffrin, J. Cook, ``Short-term forgetting of item and order information'', Journal of Verbal Learning and Verbal Behavior, Vol. 17, pp. 189~-~218, 1978.}




\bibitem{Estes} {W.K. Estes, A.W. Melton, E. Martin,  ``An associative basis for coding and organization in memory'', Coding processes in human memory (pp. 161-190). Washington, 1972.}




\bibitem{Hanlon} {A.G. Hanlon, ``Content-addressable and associative memory systems'', IEEE Transactions on Electronic Computer, Vol. EC-15, pp. 509~-~521, 1966.}






\bibitem{Katajainen} {J. Katajainen, T. Pasanen, ``Stable minimum space partitioning in linear time'', BIT Numerical Mathematics, Vol. 32, pp. 580~-~585, 1992.}
\bibitem{Salowe} {J. Salowe, W. Steiger, ``Simplifed stable merging tasks'', Journal of Algorithms, Vol. 8, pp. 557~-~571, 1987.}
\bibitem{Franceschini_1} {G. Franceschini, S. Muthukrishnan, M. Patrascu, ``Radix sorting with no extra space'', ESA'07 Proc. 15th annual European conference on Algorithms, pp. 194~-~205, 2007.}

\bibitem{Belazzougui} {D. Belazzougui, P. Boldi, R. Pagh, S. Vigna, ``Monotone Minimal Perfect Hashing: Searching a Sorted Table with $\mathcal{O}( 1 )$ Accesses'', SODA '09 Proc. of the twentieth Annual ACM-SIAM Symposium on Discrete Algorithms , pp. 785-794.}
\bibitem{rosen:discrete_math_handbook}{HK Rosen, Handbook of Discrete and Combinatorial Mathematics, CRC Press, 2000.}


  


\end{thebibliography}


\end{document}